# Web Service Interface for Data Collection

Ruchika Thukral
Department of Computer Science
University of Delhi
Delhi-110007, India
ruchikathukral2203@gmail.com

Anita Goel
Department of Computer Science
University of Delhi, Dyal Singh College
New Delhi –110003, India
agoel@dsc.du.ac.in

*Abstract*—Data collection is a key component of an information system. The widespread penetration of ICT tools in organizations and institutions has resulted in a shift in the way the data is collected. Data may be collected in printed-form, by e-mails, on a compact disk, or, by direct upload on the management information system. Since web services are platform-independent, it can access data stored in the XML format from any platform. In this paper, we present an interface which uses web services for data collection. It requires interaction between a web service deployed for the purposes of data collection, and the web address where the data is stored. Our interface requires that the web service has pre-knowledge of the address from where the data is to be collected. Also, the data to be accessed must be stored in XML format. Since our interface uses computer-supported interaction on both sides, it eases the task of regular and ongoing data collection. We apply our framework to the Education Management Information System, which collects data from schools spread across the country.

**Keywords-Education Management Information System, Web Services, XML**

I. INTRODUCTION

Data collection is the way to gather raw data to address important issues. The collected data is the evidence of reliability and relevancy. The collected data is integrated and analyzed to provide relevant information, for use during the decision making process. Management Information System (MIS) is designed and developed with an aim of collecting, organizing, integrating and analyzing the data gathered from different sources like divisions, departments and levels of an organization. During data collection, there is a need to focus on the quality, accuracy and reliability of data [18], since policies and decisions are based on the collected data. Data collection is a crucial task for an MIS.

Data collection requires a questionnaire or a form for which the data is collected. Traditionally, data is collected using approaches like interviews, face to face interactions, telephonic conversation and quantitative approaches which include mass collection of data using fax, paper-copy etc. With the introduction of computers, there has been a drastic change in the method of data collection – computerized forms are designed and data is collected via e-mails, sent on CD's, etc. Web-based surveys and e-mail have been used for sending and receiving questionnaires or data capture forms, to collect the data [13].

The collection of data is a time consuming process due to geographical disparities. Data capture formats are quite lengthy and time taking. Online data entry is tedious and increases the cost of communication over the network. Moreover, heterogeneous systems do not facilitate real time digital data transfer.

The purpose of this paper is to implement web service interface in data collection process. Web Service interface can be used to solve the problems occurring in the data collection phase of any Management Information System or research programs. Web service provides an interface to collect data from heterogeneous environments [20]. Recoding of the running system is also not required [10]. Web service interoperable feature allows to access files across different private networks also. Web service communicates in XML which is standard language over the network and any language can interact with this language. Web service can also leverage on existing protocols like HTTP, FTP etc.

It was a challenging task to collect data from all the school by National administrators timely. This project was assigned to NUEPA (National University of Educational Planning and administration). It started with the design of DCF (Data Capture format) published and disseminated to all the schools following proper channel. Channel begins from National administration —> State —>Districts —>Blocks —> Clusters —> schools. DCF disseminates through this channel and collects back through the same channel in reverse. Data entry in the system is possible at District level where data aggregates from all the blocks of a district. Data entry takes long time due to the collection of many forms at same place which delays the aggregation of data at national level. Sometimes there are wrong entries and rectification process is also time consuming as it cannot be done without the permission of National authorities because data get freeze once it is submitted. We apply the web service interface to collect data from schools.

In this paper, Section 2 focuses on the data collection methods. Section 3 discusses the challenges faced during data collection. Section 4 discusses the use of web service interface in data collection. Section 5 describes the Web service: concept, benefits and features. Section 6 illustrates a case study on Web Service interface on EMIS. Section 7 is discussion on the propose system with web service interface. Section 8 is about related work and section 9 states the conclusion.

## II. DATA COLLECTION METHODS

Data collection begins with designing of questionnaire which can be open-ended for mass communication or close-ended for an organization. Designing the questionnaire and interviewing people personally to collect the data is the traditional method. In mass collection sending the forms by post mail, requesting them to fill in accurate information and send it back by post, is a time consuming process [14]. With the introduction of computer during 1990's there was a drastic change from paper bound collection system to computerized forms which has been sent recorded in disks. Other side used to download the form fills in and send it back in the form of printed forms or digital form saving in secondary storage device via post. Telephone based methods were used to collect data which were assisted by fax machines [24]. With the introduction of Internet, web or e-mail surveys have been used for sending and receiving questionnaires or data capture forms to collect the data. Data can be filled on the web site of data collectors where different formats are provided. EDI's have been used to collect data over cross organizations but there are terms and conditions for communication and should be on homogenous environments.

## III. CHALLENGES IN DATA COLLECTION

Data filled in questionnaire or formats should reach in time but many times due to communication system it gets delayed which affects the decision making process. Data filled in forms is sometimes not as per the instructions given to fill the form. Default entries were not filled in properly. Personnel visit is required for collecting the accurate and reliable data [3]. After the introduction of ICT, for filling in forms offline or online, extra cost was spend to set up hardware and software because data transfer is possible between homogeneous environments [2]. Special trainings have been given to enter in the data in the format. After receiving the data, it has to be entered in the database for analysis. There are high chances of wrong entries which results in prolonged decisions. Cost of using internet based system is higher as longer data capture format takes long time to fill in online and cost of sending is also high. Unwillingness of institutes to change code for communication as data can only be collected from same platform and coding should be in the same language [10]. Computer transaction is not possible between two different operating system like Windows or Unix and between different languages like Visual basic supported by Windows and Sun JAVA supported by LINUX or UNIX. Cost of restructuring the system from scratch is quite high and institutes show their reluctance to bear such amounts. Planning does not mean decision of only one department. Many departments collectively frame policies for any organization. But data in heterogeneous platforms creates sharing problems [10][1].

## IV. WHAT IS WEB SERVICE?

A Web service (WS) is a *unit of functionality* exposing an XML interface, describing a WS in terms of the messages it receives and generates. It is a *self-describing*, *self-contained* entity that can be *registered* with a registry and located by the potential users. Web service is capable of being defined, described and discovered that means it is accessible and based on 'internet oriented standard based Interfaces'. WS is similar to services as traditional middleware. Not only they are running but they can be described and advertised so that it is possible for clients to bind and interact with then. WS are components that can be integrated into more complex distributed application. Web services immerged as an easy to use technology to support the concept of software as a service. They provide developers with XML-based methods to access and integrate services available over the Internet and utilize them as part of their own applications. The concept further advanced with time to result in a larger, more sophisticated framework under the service-oriented architecture (SOA). Web Service Standards are accepted by most of the organizations.

Web Service is based on new set of standards that allows it to be interoperable and technically provides possibilities to interact between heterogeneous environments. A Web Service interface complies with the following standards: XML (eXtensible Markup Language) document is used for data input and output. XML is used for this purpose, both because it is a widely adopted and commonly accepted standard language. Web service can leverage over existing file transfer protocols like HTTP of FTP. SOAP (Simple Object Access Protocol) is the standard specifying how XML documents are exchanged over HTTP. WSDL (Web Services Description Language) is used to provide information about the web service so that a user can easily identify it and use the functionalities of web service. UDDI (Universal Description, Discovery and Integration) is used to register the Web Service. Web services are registered under the UDDI from where any user can search for the related web services and use the best suited one [7][8][13][15][19]. Feature of web service also support system of data collection as Standard protocols of IT industry like XML messaging, SOAP, WSDL are used by web services which makes it more acceptable for communication between different organization. Cost of communication has lowered down as compared to EDI or other data capture format which take long time to fill in.

Web service gives flexibility to use the communication means present on network like services can also be implemented using FTP or HTTP or SOAP. Web Services are loosely coupled application and can be used by applications developed in any technologies. Developing time of web service is less as they are self describing. Applications of other organizations uses web service which gives its own description to user as web service description language. Web services are interoperable as they work outside the private networks [18]. Programmers are allowed to use language of their preference and operating system and with web service they can communicate between different applications. Web service provides zero coding deployment of existing services which makes easy to reuse web services component in different ways.

V. WEB SERVICE- INTERFACE FOR DATA COLLECTION

Due to non standard interface, such problems occur to manage any Management Information System or in research projects. These problems can be resolved by using web service as a programming interface to collect data from heterogeneous environments. Web service provides a platform where one application can exchange information with another application, or machine to machine communication is possible irrespective of the programming language and operating system [15]. Web service provides interoperability to access files across different private networks i.e. loose coupling. Web service is based on standard like XML, SOAP and WSDL which is highly accepted by most of the organizations. It interoperates using messages that does not require continuous network connection which cut the cost as compared to traditional method of transactions. Web service allows sharing the data and functionalities, which solves the problem of recoding the application again time and money saving [11][21] (Figure 1)

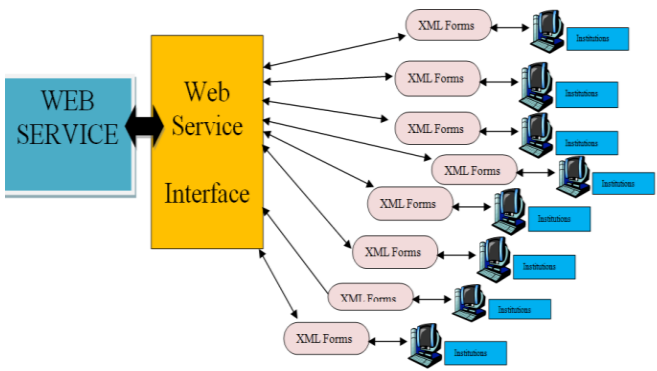

Figure 1. Working of Proposed System using Web Service Interface

VI. CASE STUDY: SCHOOL EMIS IN INDIA

Educational Management Information Systems (EMIS) is a system to store data from educational institutes, retrieve, and process and analyzed by decision makers to plans policies for educational system of a country [22]. Data from educational institutes are collected and national administrators use the data for further analysis [8]. Education Management Information System (EMIS) collects data as raw from secondary school, stores, analyzes and provides information about present scenario of educational institutes of a country [9][14]. In India education policies are framed by Ministry of Human Resource Development. Statistical data of school education is collected by three agencies a) Planning, Monitoring and statistics division, Department of Education, MHRD; b) University Grant Commission on Higher Education; and c) National Council Of Educational Research and Training (NCERT) on occasional basis [5][6][11]. Earlier data from schools are manually collected and recorded then with DISE in 2000-01 efforts were made to computerize the Management Information System. By the end of 2003, DISE extended to about 460 out of 593 districts of the country. These districts are spread over 18 DPEP states. The Government of India decided that the manual collection of information system would gradually be replaced by the DISE and the statistics generated by it will be accorded the status of the Official Statistics. Two projects DISE and SEMIS run by NUEPA to collect the data from educational institutes are undertaken by MHRD. DISE is connected with Primary and Elementary education of the country where data is collected from 13,00,000 schools. Secondary Education data is collected in program named as SEMIS (Secondary Education Management Information System) from 2,00,000 schools of the country [5][6][8]. All these schools are recognized irrespective of aided or non-aided by Government. The dissemination of Data Capture Format in DISE starts from National level to State then districts to blocks, blocks to cluster and finally to school

School Head Masters have to fill these forms according to the instructions given by the planning council. After filling it has to be sending back to clusters. 10 to 15 schools are there in each cluster. At block level DCF are collected and send to districts educational project officers (DO) with the help of secondary storage devices. It is then entered in the offline computerized DCF and forwarded to state educational project officers (Director of Education). Aggregation of data is done at State level and State level Reports are generated to publish the data of schools. SEMIS is online EMIS where DCF is uploaded on web site. At District level DCF are downloaded and distributed to blocks further clusters and then to schools from proper channel *(see Figure 2)*.

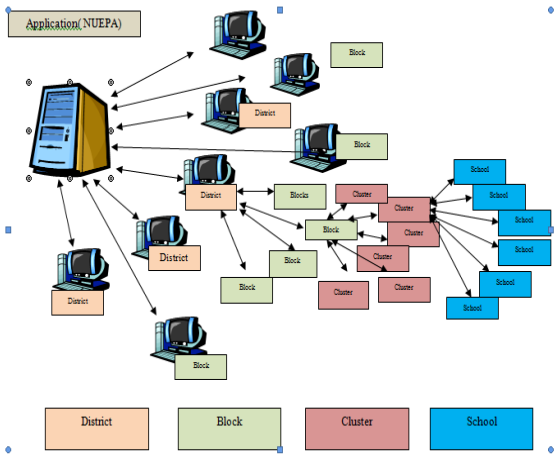

Figure 2. Working of SEMIS

Headmasters have to fill in the form and collected back by cluster and then to blocks. Blocks handover the data in paper form to districts, or, in digital form if they have computerized systems. At District level data is entered online and gets freeze by National administrators. This displays the report card of every state/districts and then can be downloaded by States for their record. To measure the EMIS success the data should be [10][13]:

- Reliable
- Timely collected
- Shared by other departments
- Effectively used by policy makers

There are few drawbacks of this system. Dissemination and collection has long way from National administrators to school and then from schools to national level. Due to geographical diversions and more than 600 districts to be set up with computerized environment, cost of implementing the system is high. Hardware and software cost increase the budget of implementing educational management system. Special trainings have been given to enter the data in DCF online forms with all the default values and codes keeping in mind. Also trainings have been given to headmasters to fill in the printed forms which increase the budget of the project. Cost of transferring the files in SEMIS is high as the 23 pages long form takes time to download and upload. Filling in the form online, adds internet cost. Besides computerized on line system, EMIS has also got few limitations. Data filled in once cannot be changed if submitted. Only National level administrators can give the permission to rectify the data. Data Capture Format can be downloaded at District level and then disseminated to different schools in channel discussed above. Filled in forms are entered at District levels only. So, time taken in this dissemination and collection is long which delays the processing. Data entry is done by officers at district levels or sometimes at block level. Sometimes data entered is wrong and rectification takes time which again causes the delays [8][14][23]. Hardware and software problems delay the collection process which directly causes the delays in decision processes. Real time filling of forms from schools is not possible due to heterogeneous platforms and application in different languages. To have homogeneous systems and changing the coding is not accepted by the schools. Annual budgeting and policy making process of the country depends on many departments. Data needs to be shared but different hardware and software configuration barricade them. Data integration and sharing of data between different departments of same organization is lso not possible because of hardware and software disparities.

VII. PROPOSAL TO USE THE WEB SERVICE INTERFACE

Integration of data by using Web Services interface can simplify the current EMIS. EMIS can send XML forms directly to the schools accounts and can be downloaded by the school personnel (Figure 3). Web Service communicates in XML, which is a standard web communication language and easy to understand. DCF is designed in XML forms using open office software which is free downloadable and can be used on any operating system.

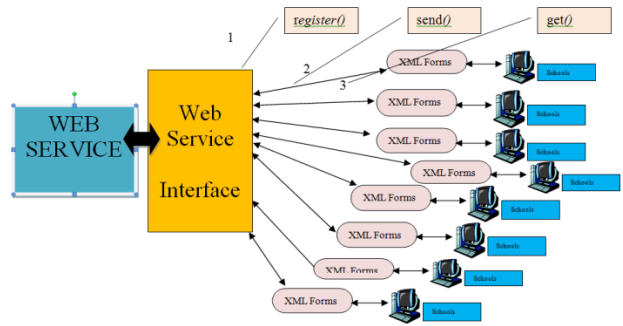

Figure 3. Working of Proposed System using Web Service Interface

Without redesigning the current system of school, DCF can be used and filled in by head masters offline. They are required to save XML form as XML file and upload to the school web site. Now web service which is the unit of functionality of EMIS designed in JAVA Eclipse using Apache Tomcat as server will pick the XML file with School ID. JAVA methods are used to design the web service interface where -

- o *register()* method is used to register the web site URL's of schools;
- o *send()* method is used to send XML form to registered URL's accounts;

- and *get()* method is used to get back filled in XML form from registered URL's of school;

Web service interface will pull the file on weekly, fortnightly or monthly basis. This is much faster than the current system of data collection. Data aggregation can be done at any level whether District, State or National. Advantages of using web service interface over current EMIS are that the web service can communicate with any application designed in any language (VB, JAVA) supported by any operating system (Windows, UNIX, LINUX). Web Service use web standards like XML, SOAP, WSDL which are universally accepted. Advantages to implement the system are - cost of communication is less, data can be shared within the different departments of an organization or between different organizations, real time communication is possible, data entries are verified-less chances of mistakes, rectifying of mistakes is much faster. So, it can be said that system with web service interface will improve the data collection process by increasing reliability, accuracy, cost efficiency and faster.

## VIII. RELATED WORK

Data collection is the process of collecting data from different divisions of an organization, many organizations, mass collection of data for research projects etc. We have many ways to collect the data. From traditional paper based data collection method to online submission of data, data collection has changed many folds in the speed, accuracy, reliability and completeness [3][12][18]. Challenges faced during the data collection are time delays, increase in cost due to printing of forms, dissemination of forms, collection, rectifications, trainings to use on line forms for valid entries, data sharing and integration and recording [1][13][10].

Web service interface can be used to provide a platform for heterogeneous interactions, interoperability, timely and reliable data collection because it is based on set of standards like XML, WSDL and SOAP [4][7][8][13][15][19]. For case study EMIS and SMIS have been taken to implement Web Service interface [5][6][8][9][14]. Use of web service will increase the cost efficiency and reliability of data. Data can be aggregated and shared among departments without recoding the entire system from scratch [10][18][13].

## IX. CONCLUSION

The purpose of this paper is to find solution for the limitations to collect data from the institutes all over the country. Due to geographical disparities, personal interviews or group meetings are not possible. From pen-paper data collection methods to computerized systems of data collection, there were many challenges faced by data collectors. For any Research project or MIS, data collection is very important task as the information generating process is based on timely and reliable data. Many problems like interoperability between networks, communication between heterogeneous environments, understanding different languages, cost of redesigning of entire system, cost on training to understand the complex and lengthy questionnaire etc can be solved by using web service interface. This paper proposes a web service interface which sends and receives messages in XML standards of communication over the network. Cost of designing such system is low as there is no need to redesign the complete system from scratch. Also communication cost is less because web service sends and receives discrete message packets which do not require continuous Internet connection.

REFERENCES

[1] Ana-Ramona Lupu, Razvan Bologa, Gheorghe Sabau, Mihaela Muntean "Integrated Information Systems in Higher Education" WSEAS TRANSACTON on COMPUTERS ISSN: 1109-2750 MAY 2008

[2] Andrew p., Ciganek, marc n. Haines, william d. Haseman "service-oriented architecture adoption: key factors and approaches" journal of information technology management volume xx, number 3, 2009

[3] Annette Jäckle, Caroline Roberts, Peter Lynn "Assessing the Effect of Data Collection Mode on Measurement" ISER Working Paper Series Feb. 2008.

[4] Ali Arsanjani, Brent Hailpern, Joanne Martin, Peri L. Tarr "IBM Research Report Web Services: Promises and Compromises "RC22494 (W0206-107) June 20, 2002.

[5] Arun C. Mehta " DISE II: HE-MIS Higher Education Management Information System" ACM/HE-MIS/Sept 14, 2007

[6] Arun C. Mehta " Status of Educational Management Information System in Iraq and Suggestions for Improvement" Workshop on EMIS and Educational Statistics for the Officers of Iraq, Dead Sea, Amman, Jordan, 03 to 15 April, 2006

[7] Ashok Kumar ICT in Delhi School Education System Published by University Press, 2008.

[8] Australian Government " Hig he r e ducation inf o r matio n management s ys te m (he ims ) he ims web s ervices interface technical s p e c if ic atio n chessn f unctions" , Commonwealth of Australia 2008

[9] Azlinah Mohamed, Nik Abdullah Nik Abdul Kadir, Yap May-Lin, Shuzlina Abdul Rahman, and Noor Habibah Arshad, "Data completeness analysis in the Malaysian Educational Management Information System" (IJEDICT), 2009, Vol. 5, Issue 2, pp. 106-122

[10] California Department of Education " CALIFORNIA SPECIAL EDUCATION MANAGEMENT INFORMATION SYSTEM (CASEMIS) Technical Assistance Guide December 2010.

[11] "Educational Management Information System in India" http://www.educationforallinindia.com/page3.html

[12] Francois, Mike, Andrew Miller "Using Para Data To Actively Manage Data CollectionSurvey Process" JSM 2008.

[13] John DIXIE and Kenneth ANDERSON "Harnessing New Technologies To Improve Data Collection From Local Education Authorities And Schools In England" http/

*epp.eurostat.ec.europa.eu/cache/ITY_PUBLIC/NTTS2001/63.pdf*

[14] Haiyan Hua and Jon Herstein "Education Management Information System (EMIS): Integrated Data and Information Systems and Their Implications In Educational Management" Annual Conference of Comparative and International Education Society March 2003

[15] Mark Hansen, Stuart Madnick, Michael Siegel "Data Integration using Web Services" Working Paper 4406-02 CISL 2002-14 May 2002

[16] Marcus Powell, "Rethinking Education Management Information Systems: Lessons from and Options for Less Developed Countries" WORKING PAPER NO. 6, 2006

[17] Marie Lall," The Challenges for India's Education System" ASIA PROGRAMME ASP BP 05/03 APRIL 2005

[18] McDonald, Heath and Adam, Stewart "A comparison of online and postal data collection methods in marketing research, *Marketing intelligence & planning*" vol. 21, no. 2, pp. 85-95 2003.

[19] Nikolaos v. Karadimas ,nikolaos p. Papastamatiou "digital collections serving as educational repositories" i.j. of simulation vol. 9 no 1 issn 1473-804x online, 1473-8031

[20] Ruchika and Anita Goel "Web Services for Education Management" 4th International Conference on Data Management Institute of Management Technology Ghaziabad, India(in press) November 2011.

[21] Shen, W.; Hao, Q.; Mak, H.; Neelamkavil, J.; Xie, H.; Dickinson, J.K.; Thomas, J.R. Pardasani, A.; Xue, H "Systems integration and collaboration in architecture, engineering, construction and facilities management: a review" Advanced Engineering Informatics, 24, (2), September 01, 2009

[22] Shooebridge, JInfoDev study #1253 "Education Management Information System: Case Studies and Lessons Learnt, Case Study of Nigeria" (2006)

[23] Tom Cassidy "Education Management Information System (EMIS) Development in Latin America and the Caribbean: Lessons and Challenges" January 2005.

[24] William L. Nicholls II, (Retired) and Thomas L. Mesenbourg Jr., U.S. Census Bureau, Stephen H. Andrews, "Use Of New Data Collection Methods In Establishment Surveys" *www.amstat.org/meetings/ices/2000/proceedings/S50.pdf*